\documentclass{SCGE}

\begin{document}

\begin{picture}(0,0){\rm
\put(0,-20){\makebox[160truemm][l]{\bf {\sanhao\raisebox{2pt}{.}}
Article  {\sanhao\raisebox{1.5pt}{.}}}}}
\put(0,-34){\jiuwuhao {\textcolor[rgb]{0.5,0.5,0.5}{\sf 
}}}
\end{picture}

\def\bm{\boldsymbol}

\def\dl{\displaystyle}
\def\du{\end{document}}
\def\d{{\rm d}}
\def\e{{\rm e}}
\def\i{{\rm i}}
\def\pi{{\uppi}}

\Year{?} %
\Month{??} %
\Vol{?} 
\No{?} 
\BeginPage{1} 
\AuthorMark{{\rm Junjie Qi}, et al.}  
\AuthorMarkCite{{\rm Junjie Qi}, et al. } 
\DOI{??} 
\ArtNo{??}

\title[Effective spin dephasing mechanism in confined two-dimensional topological insulators]
{Effective spin dephasing mechanism in confined two-dimensional topological insulators}

\author[1]{Junjie Qi}{}
\author[2]{Haiwen Liu}{haiwen.liu@bnu.edu.cn}
\author[3]{Hua Jiang}{}
\author[4,5]{X. C. Xie}{xcxie@pku.edu.cn}

\address[{\rm1}]{Institute of Physics, Chinese Academy of Sciences, Beijing 100190, China;}
\address[{\rm2}]{Center for Advanced Quantum Studies, Department of Physics, Beijing Normal University, Beijing, 100875, China;}
\address[{\rm3}]{College of Physics, Optoelectronics and Energy, Soochow University, Suzhou 215006, China}
\address[{\rm4}]{International Center for Quantum Materials and School of Physics, Peking University, Beijing 100871, China;}
\address[{\rm5}]{Collaborative Innovation Center of Quantum Matter, Beijing, China;}

\maketitle

\begin{center}
\rule{16.5cm}{0.4pt}
\parbox{16.5cm}
{\begin{abstract}  A Kramers pair of helical edge states in quantum spin Hall effect (QSHE) is robust against normal dephasing but not robust to spin dephasing. In our work, we provide an effective spin dephasing mechanism in the puddles of two-dimensional (2D) QSHE, which is simulated as quantum dots modeled by 2D massive Dirac Hamiltonian. We demonstrate that the spin dephasing effect can originate from the combination of the Rashba spin-orbit coupling and electron-phonon interaction, which gives rise to inelastic backscattering in edge states within the topological insulator quantum dots, although the time-reversal symmetry is preserved throughout. Finally, we discuss the tunneling between extended helical edge states and local edge states in the QSH quantum dots, which leads to backscattering in the extended edge states. These results can explain the more robust edge transport in InAs/GaSb QSH systems.
\end{abstract}}
\end{center}\vspace*{-0.6cm}

\begin{center}
\parbox{16.5cm}
{\bf\jiuhao Quantum spin Hall effect, bound helical states, spin dephasing, Rashba spin-orbit coupling, electron-phonon interaction}
\end{center}

\begin{center}
{}
\CITA    
\end{center}

\textwidth=178truemm \textheight=236truemm

\wuhao\vspace*{1.5mm}

\begin{multicols}{2}

\renewcommand{\baselinestretch}{1.08} \baselineskip 12.2pt\parindent=10.8pt

\renewcommand{\thefootnote}

\section{Introduction}\label{sec:intro}

Recently the time reversal symmetry protected topological insulators (TIs) has attracted great interests \cite{RevModPhys.82.3045,RevModPhys.83.1057}. The two dimensional TIs-quantum spin Hall (QSH) systems have been theoretically predicted in many systems \cite{PhysRevLett.95.226801,PhysRevLett.95.146802,Bernevig,PhysRevLett.100.236601,YaoYG,WengHM} and experimentally observed in heterostucture systems, e.g., HgTe/CdTe quantum wells(QWs) \cite{Konig02112007} and in InAs/GaSb QWs \cite{PhysRevLett.100.236601}.  The QSH system has insulating bulk states with an energy gap, and simultaneously possesses metallic helical edge states on the boundary which are protected by the topological property of bulk states \cite{RevModPhys.82.3045,RevModPhys.83.1057}. This peculiar property is characterized by a $\mathbb{Z}_{2}$ topological invariant \cite{PhysRevLett.95.146802}. Specifically, when time-reversal symmetry (TRS) is preserved, two counter-propagating edge electrons form a Kramers pair which is immune to non-magnetic impurity backscattering. The two-dimensional topological phases exhibit potentials for device manufacture due to the existence of non-dissipative edge states.

In realistic systems, there exist two categories of dephasing process: normal dephasing and spin dephasing. Normal dephasing, caused by electron-phonon interactions or the electron-electron interactions etc., contributes to phase relaxation but does not flip spin of electrons \cite{CHAKRAVARTY1986193,PhysRevA.41.3436}. On the other hand, spin dephasing process affects both phase relaxation and spin flip, such as the dephasing caused by magnetic impurities \cite{RevModPhys.76.323,ActaPhysSlovaca}. Previous studies have shown that the Kramers pair of helical edge states is robust against normal dephasing but fragile with spin dephasing \cite{PhysRevLett.103.036803}. For spin-momentum locked helical surface states of 3D topological insulators, the combination of normal dephasing and impurity scattering can also cause backscattering \cite{dephasing2014}. Experimentally, transport measurements in InAs/GaSb QWs demonstrate that the spin dephasing can be the dominant edge scattering process in this system \cite{Du2014}. Moreover, it has been shown that spin dephasing can originate from the combination effect of spin-orbit coupling(SOC) and normal dephasing effect \cite{PhysRevLett.83.1211}.  Thus, for the QSH systems, the simultaneous existences of Rashba SOC  and normal dephasing may also lead to an effective spin dephasing, and further give rise to inelastic backscattering in the helical edge states. Specifically, the Rashba SOC term can cause spin flip and usually have two different origins: (i) from the axial symmetry breaking, when a top gate is applied on HgTe/CdTe QWs \cite{1367-2630-12-6-065012}; (ii) from the structure inversion asymmetry (SIA) in InAs/GaSb QWs due to the unique band alignment in the QWs \cite{PhysRevLett.100.236601}. Nevertheless, the mere existence of Rashba SOC cannot cause backscattering, due to the protection of TRS . Several papers have reported two kinds of mechanism on inelastic backscattering of helical edge states caused by Rashba SOC and normal dephasing. For normal dephasing caused by electron-electron interaction, the momentum linear $k$-order Rashba SOC can give rise to backscattering \cite{PhysRevLett.108.156402,PhysRevLett.110.216402}. In the relatively high temperature region, the electron-phonon interaction may play more important role in the normal dephasing process. However, in extended helical edge states, due to the elimination of Feynman diagrams, the $k$-order Rashba SOC and normal dephasing (from electron-phonon interaction) cannot lead to backscattering, and only the $k^3$-order Rashba SOC term survives and contributes to the backscattering process \cite{PhysRevLett.108.086602}. But the linear $k$-order Rashba SOC is more general and larger than the $k^{3}$-order one around the $\varGamma$ point, and the elimination of Feynman diagrams may not happen in the confined system.  Thus, we are interested in (i) whether or not spin dephasing and inelastic backscattering exist in a QSH quantum dot with the combination of the $k$-order Rashba SOC and electron-phonon interaction; (ii) if it exists, how does this mechanism work and further influence the inelastic backscattering properties of extended helical edge states?

In our paper, we find that, in confined QSH quantum dots, the combination of $k$-order Rashba SOC and electron-phonon interaction results in an effective spin dephasing mechanism. For the local helical edge states around the boundary of QSH quantum dot, the contribution of $k$-order Rashba SOC cannot be eliminated, compared to the case of extended helical edge states. The QSH quantum dots, or metallic puddles of QSH, usually exist in doped semiconductor hetero-structures with the inhomogeneity, because the donors and acceptors are inevitably introduced by sample growth \cite{PhysRevX.3.021003,Konig02112007,Roth17072009}. We simulate the QSH quantum dots (QDs) modeled by the massive Dirac Hamiltonian as shown in Figure 1(a), and consider the topological boundary with mass term changing sign at the radial boundary, which maintains the existence of the helical bound states around the boundary. The confined system can also be labeled by a $\mathbb{Z}_2$ quantum number since the existing of the helical bound states. Moreover, we take into account the linear $k$-order Rashba SOC and electron-phonon in the QD and calculated the spin-dephasing rate between helical edge states in the QSH QDs. The results demonstrate that inelastic backscattering is unavoidable in the QSH QD with the combination of the linear $k$-order Rashba SOC and electron-phonon interaction. Finally, we discuss the tunneling between  extended helical edge states and local edge states in the QSH QD.

\begin{figure}[H]
\centering
\includegraphics[scale=0.4]{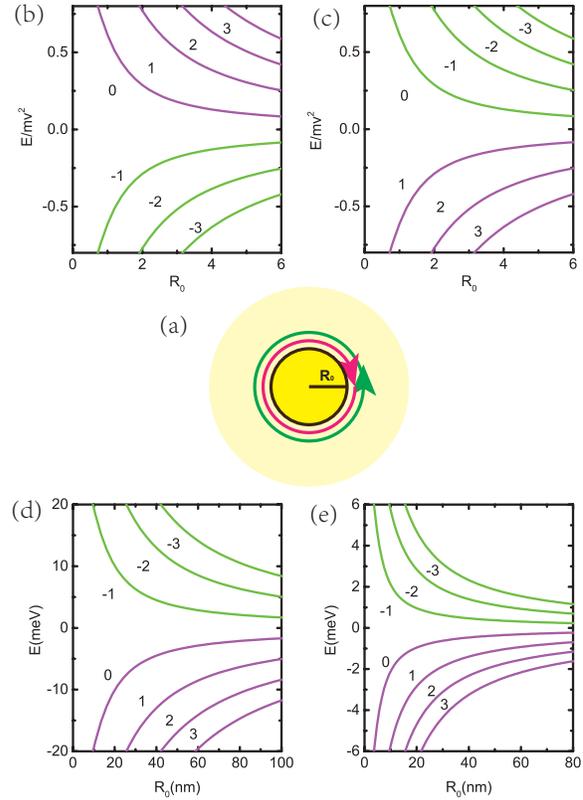}
\caption{(Color online) (a) Schematic plot of a two-dimension QSH QD. Electrons with different spin directions have opposite angular momenta and run oppositely along the edge. The disk represents a QSH QD which is described by the two-dimensional massive Dirac Hamiltonian. The sign of mass term in Dirac Hamiltonian differs around the radial boundary $R_{0}$ in the QSH QD. And there exists a pair of helical states circulating along the boundary. (b),(d) and (e) Energy levels of spin-up part of topological boundary of a QD versus radius $R_0$. Different curves represent different rotation directions. In (b) and (c), we set $m=v=\hbar=1$;(b) and (c) are the energy levels of spin-up and spin-down parts respectively; in (d), $mv^2=-24.8$ meV,$\hbar v=333.6$ meV.nm \cite{Liuchaoxing}; in (e), $mv^2=-7.8$ meV,$\hbar v=37$ meV.nm \cite{Liuchaoxing}.} 
\label{fig:figure1}
\end{figure}

The paper is organized as follows. In sect. 2, we employ the massive Dirac Hamiltonian to describe a QSH QD. In sect. 3, we obtain the helical bound solution circulating the radial boundary. In sect. 4, we calculate the spin dephasing relaxation time. Furthermore, we discuss inelastic backscattering results and the tunneling between extended helical edge states and a QD. Finally, we end our analysis with a conclusion in sect. 5.

\vspace*{-1mm}

\section{Model}\vspace*{-1mm}

We start by considering a two-dimension QSH QD shown in Figure 1(a), which can be described by the massive Dirac Hamiltonian. In the following, we give out the $2\times2$ matrix Hamiltonian for each spin index. Here the Hamiltonian for spin up part is represented by $h(k)$ and the spin down part $h^*(-k)$ is the time reversal symmetry counterpart of $h(k)$.

\begin{equation}
\begin{aligned}
h(k)&=\left(
\begin{matrix}
M(\vec{r})v^2& v p_{-}\\
v p_{+}&-M(\vec{r})v^2
\end{matrix}
\right)
\end{aligned},
\end{equation}

\begin{equation}
\begin{aligned}
h^{*}(-k)&=\left(
\begin{matrix}
M(\vec{r})v^{2}& -v p_{+}\\
-v p_{-}&-M(\vec{r})v^{2}
\end{matrix}
\right)
\end{aligned},
\end{equation}

and

\begin{equation}
M(\vec{r})=\left\{
\begin{array}{rcl}
m_{1}      &      & {r>R_{0}}\\
-m_{2}    &      & {r<R_{0}}
\end{array}\right.
\end{equation}

where $m_{1}$ and $m_{2}$ are mass terms possessing positive values and $v$ is the Fermi velocity. As illustrated in the following part, the term $m(\vec{r})$ leads to a bound solution at $r=R_{0}$ when the mass sign is different inside and outside the boundary $r=R_{0}$. Previous studies have shown that soliton solution exists around the boundary \cite{Jackiw,ZhangF}. For this reason, we call the mass sign change at $r=R_{0}$ a topological boundary. In confined model with central symmetry, it's convenient to use the cylindrical coordinate $(x,y,z)=(rcos\theta,rsin\theta,z)$, and the Dirac Hamiltonian can be deduced as:

\begin{equation}
p_{\pm}=p_{x}\pm i p_{y}=- i\hbar e^{\pm  i \theta}(\frac{\partial}{\partial r} \pm i\frac{1}{r}\frac{\partial}{\partial\text{\ensuremath{\theta}}}).
\end{equation}

Due to the TRS, the bound solution is a pair of helical states circulating around the disk, as shown in Figure 1(a). Since the $z$-component of the total angular momentum $j_{z}=-i\hbar\partial_{\theta}+(\hbar/2)\sigma_{z}$ can be a good quantum number \cite{PhysRevB.84.035307}, we would like to adopt $\psi_{n\sigma}=\vert n\sigma\rangle$ to label the helical state in the QSH QD. $n$ is the angular momentum in $z$ direction which is labeled by an integer $n\in\{0,\pm1,\pm2,...\}$.

A Rashba SOC term $h_{R}$ originates from the axial symmetry breaking \cite{PhysRevLett.104.256804}. More importantly, the electric potential at the boundary of puddles causes a perpendicular electric field and gives rise to the Rashba SOC.
\begin{equation}
\begin{aligned}
h_{R}=\left(
\begin{matrix}
\alpha e^{-i\theta}(\frac{\partial}{\partial r} - i\frac{1}{r}\frac{\partial}{\partial\text{\ensuremath{\theta}}}) &0\\
0 &0
\end{matrix}
\right)
\end{aligned},
\end{equation}

where $\alpha$ is the Rashba SOC strength. In our calculation, we treat Rashba SOC term as perturbation. The total Hamiltonian of our setup has a $4\times4$ matrix structure which read

\begin{equation}
\begin{aligned}
H=\left(
\begin{matrix}
h(k)& h_{R}\\
h_{R}^{\dagger}&h^{*}(-k)\\
\end{matrix}
\right)
\end{aligned}.
\end{equation}

The wave-function has the form $\Psi=(\psi_{n,\uparrow},\psi_{n',\downarrow})^{T}$. These two spin-generate wave functions $\psi_{n,\uparrow}$ and $\psi_{n',\downarrow}$ are spinors which are related to spin-up and spin-down states, respectively.
Another key element is the electron-phonon interaction introduced by
a classical description. Given the inhomogeneities at the boundary of the puddles, the frequency of optical (local) phonons should have a broadening spectra around the central energy. The details of phonons are illustrated in sect. 4.

\section{Helical bound states}
In this section, we solve the bound solution of a QSH QD around the topological boundary. We begin with solving the spin-up part $h(k)$ outside the $R_{0}$. When $r>R_{0}$, the Hamiltonian is

\begin{equation}
\begin{aligned}
h(k)&=\left(
\begin{matrix}
m_{1}v^{2}& -i\hbar v e^{-i\theta}(\frac{\partial}{\partial r} - i\frac{1}{r}\frac{\partial}{\partial\text{\ensuremath{\theta}}})\\
 -i\hbar v e^{i\theta}(\frac{\partial}{\partial r} + i\frac{1}{r}\frac{\partial}{\partial\text{\ensuremath{\theta}}})&-m_{1}v^{2}
\end{matrix}
\right)
\end{aligned}.
\end{equation}

And we set the wave-function with the form

\begin{equation}
\begin{aligned}
\psi_{n\uparrow}&=\left(
\begin{matrix}
f_{n}(r)e^{in\theta}\\
 g_{n}(r)e^{i(n+1)\theta}
\end{matrix}
\right)
\end{aligned}.
\end{equation}

Thus, the equation is reduced that for the radial part of the wave-function

\begin{equation}
\begin{aligned}
h_{r}&=\left(
\begin{matrix}
m_{1}v^{2}& -i\hbar v (\frac{\partial}{\partial r}+\frac{n+1}{r})\\
 -i\hbar v (\frac{\partial}{\partial r}-\frac{n}{r})&-m_{1} v^{2}
\end{matrix}
\right)
\end{aligned}.
\end{equation}

We are not interested in extended bulk states here. And we only consider the bound states near the junction with the
 boundary condition $\psi_{\mid r=+\infty}=0$. The general wave function can take the form  $\psi_{n,\uparrow}=(\psi_{1},\psi_{2})^{T}e^{-\lambda_{1} r}$.  The in-gap states exists in the regime with $m_{1} v^{2}>E_{n\uparrow}$ and real $\lambda_{1}$. Then the determinant equation gives $\lambda_{1}^{2}=\frac{m_{1}^{2} v^{4}-E_{n\uparrow}^{2}}{\hbar^{2} v^{2}}$.  We choose positive $\lambda_{1}$ to satisfy the boundary condition at $r=+\infty$:

\begin{equation}
\lambda_{1}=\frac{\sqrt{m_{1}^{2} v^{4}-E_{n\uparrow}^{2}}}{\hbar v},
\end{equation}

where $m_{1}v^{2}>E_{n\uparrow}$. Conversely, for $m_{1}v^{2}<E_{n\uparrow}$, we get the pure imaginary solutions which indicates that the corresponding wave function spreads in space. The two components in the wave function satisfy

\begin{equation}
g_{n}=\frac{-i\hbar v(\partial{r}-\frac{n}{r})}{m_{1}v^{2}+E_{n\uparrow}}f_{n}.
\end{equation}

Then we substitute eq.(11) into the radial Hamiltonian $h_{r}$, and $f_{n}$ satisfies the modified Bessel equation,

\begin{equation}
 \{r^{2}\frac{\partial^{2}}{\partial r^{2}}+r\frac{\partial}{\partial{r}}-(n^2+\lambda^{2}_{1}r^{2})\}f_{n}=0,
\end{equation}

and the corresponding wave-function is

\begin{equation}
\begin{aligned}
\psi_{n\uparrow}&=A\left(
\begin{matrix}
\frac{i\hbar v\lambda_{1}}{E_{n\uparrow}-m_{1}v^{2}}K_{n}(\lambda_{1}r)e^{in\theta}\\
K_{n+1}(\lambda_{1}r)e^{i(n+1)\theta}
\end{matrix}
\right),&   &r>R_{0}
\end{aligned},
\end{equation}

where $K_{n}(\lambda_{1}r)$ is the modified Bessel function of second kind.

For the region $r<R_{0}$, we can follow the similar procedure. Then the wave function reads

\begin{equation}
\begin{aligned}
\psi_{n\uparrow}&=B\left(
\begin{matrix}
\frac{-i\hbar v\lambda_{2}}{m_{2}v^{2}+E_{n\uparrow}}I_{n}(\lambda_{2}r)e^{in\theta}\\
I_{n+1}(\lambda_{2}r)e^{i(n+1)\theta}
\end{matrix}
\right),&   &r<R_{0}
\end{aligned},
\end{equation}

where $I_{n}(\lambda_{2}r)$ is the modified Bessel function of first kind, and we have
$\lambda_{2}=\frac{\sqrt{m_{2}^{2}v^{4}-E_{n\uparrow}^{2}}}{\hbar v}$ when $m_{2}v^{2}>E_{n\uparrow}$, and A, B are the normalization factors.

Using the continuous condition at $r=R_{0}$, we arrive at the transcendental equation for $E_{n\uparrow}$ as

\begin{equation}
 \frac{i\hbar v\lambda_{1}}{E_{n\uparrow}-m_{1}v^{2}}\frac{K_{n}(\lambda_{1}R_{0})}{K_{n+1}(\lambda_{1}R_{0})}=
\frac{-i\hbar v\lambda_{2}}{m_{2}v^{2}+E_{n\uparrow}}\frac{I_{n}(\lambda_{2}R_{0})}{I_{n+1}(\lambda_{2}R_{0})}.
\end{equation}

Here we consider a simple case when $m_{1}=m_{2}=m$ and thus $\lambda_{1}=\lambda_{2}=\lambda$

\begin{equation}
\frac{E_{n\uparrow}}{mv^{2}}=\frac{K_{n+1}(\lambda R_{0})I_{n}(\lambda R_{0})-
K_{n}(\lambda R_{0})I_{n+1}(\lambda R_{0})}{K_{n+1}(\lambda R_{0})I_{n}(\lambda R_{0})
+K_{n}(\lambda R_{0})I_{n+1}(\lambda R_{0})}.
\end{equation}

The solution of the counterpart $h^{*}(-k)$ can be treated in a similar way. Cause our model preserves TRS, we can get the solution conveniently through symmetry analysis. For spin-down case, we use the angular momentum $n'\in\{0,\pm1,\pm2,...\}$ to label the wave function. The Kramers pair $\psi_{n\uparrow}$ and $\psi_{n'\downarrow}$ have the opposite angular momentum, i.e., $n'=-n$, so the transcendental equation for energies of $h^{*}(-k)$ can be written as

\begin{equation}
\frac{E_{n'\downarrow}}{mv^{2}}=\frac{K_{n'-1}(\lambda R_{0})I_{n'}(\lambda R_{0})-
K_{n'}(\lambda R_{0})I_{n'-1}(\lambda R_{0})}{K_{n'-1}(\lambda R_{0})I_{n'}(\lambda R_{0})
+K_{n'}(\lambda R_{0})I_{n'-1}(\lambda R_{0})},
\end{equation}

and the corresponding wave functions are
\begin{equation}
\begin{aligned}
\psi_{n'\downarrow}&=C\left(
\begin{matrix}
\frac{-i\hbar v\lambda_{1}}{E_{n'\downarrow}-m_{1}v^{2}}K_{n'}(\lambda_{1}r)e^{in'\theta}\\
K_{n'-1}(\lambda_{1}r)e^{i(n'-1)\theta}
\end{matrix}
\right),&   &r>R_{0}
\end{aligned},
\end{equation}

\begin{equation}
\begin{aligned}
\psi_{n'\downarrow}&=D \left(
\begin{matrix}
\frac{i\hbar v\lambda_{2}}{m_{2}v^{2}+E_{n'\downarrow}}I_{n'}(\lambda_{2}r)e^{in'\theta}\\
I_{n'-1}(\lambda_{2}r)e^{i(n'-1)\theta}
\end{matrix}
\right),&   &r<R_{0}
\end{aligned},
\end{equation}

where $\lambda_{1}=\frac{\sqrt{m_{1}^{2}v^{4}-E_{n'\downarrow}^{2}}}{\hbar v}$, $\lambda_{2}=\frac{\sqrt{m_{2}^{2}v^{4}-E_{n'\downarrow}^{2}}}{\hbar v}$ and C, D are the normalization factors.
For simplicity, we use notation $\beta=\frac{-i\hbar v\lambda}{mv^{2}+E_{n\sigma}}$ and $\beta_{1}=\frac{i\hbar v\lambda}{E_{n\sigma}-mv^{2}}$ to represent the coefficients of the wave-function in the following. The normalization factors A,B,C,D are contained in the
numerical calculations below, but not shown in the main text due to their lengthy analytic expression.

In Figures 1(b)-(e), we have shown the numerical study of energy levels of topological boundary of a QSH QD verse radius $R_{0}$. When $R_{0}$ is getting smaller, the energy level becomes much more discrete. The gap  between two arbitrary energy level is larger at small $R_{0}$. For $R_0 \rightarrow \infty$, we can deduce $E_{n\uparrow}=\frac{(n+\frac{1}{2})\hbar v}{R_0}$ of the spin-up Hamiltonian $h(k)$.
Similarly, for the time reversal symmetry counterpart $h^{*}(-k)$, we have $E_{n\downarrow}=-\frac{(n+\frac{1}{2})\hbar v}{R_0}$. For a certain angular momentum,
 the two states have opposite velocities which form a pair of helical bound states. If
 we define the wavevector $k=\frac{(n+\frac{1}{2})}{R_0}$, then the energy spectrum becomes a Dirac
 type one:$E_{n\uparrow}=\hbar v k$. The asymptotic solution can recall the extended helical edge states.

\section{Inelastic backscattering and spin dephasing mechanism}

In this part, we demonstrate how the Rashba SOC and electron-phonon interaction play a role in backscattering of local helical edge states in a TI QD. The Rashba SOC can originate from the external applied electric field or the electric field in asymmetric double quantum well. More importantly, the Rashba SOC exists at the boundary of QD and become inhomogeneous, because the potential changes fast at the boundary. Thus, we consider Rashba SOC as a perturbation effect on the helical electrons at the boundary of TI QDs. Then, we calculate the scattering matrix between opposite spin
$h_{n\uparrow n'\downarrow}^{R}=\frac{\langle n\uparrow\vert h_{R} \vert n'\downarrow\rangle}{|E_{n\uparrow}-E_{n'\downarrow}|}$. The value of the whole matrix contains two parts: the inside region $r<R_{0}$ and the outside region $r>R_{0}$.
\begin{equation}
h_{n\uparrow n'\downarrow}^{R}=h_{n\uparrow n'\downarrow}^{<}+h_{n\uparrow n'\downarrow}^{>}
\end{equation}
 In order to get the nonzero scattering values, the angular part of the integration must have $n'=n+1$ and thus
 $E_{n+1\downarrow}=-E_{n\uparrow}$. Based on our calculation, the scattering matrix only exists between two states when their angular momentum differ by one, and the scattering matrix reads
\begin{equation}
h_{n\uparrow n+1\downarrow}^{<}=-2\pi \alpha|\beta(\lambda)|^{2}\frac{
\int_{0}^{R_{0}}\lambda rI_{n}^{2}(\lambda r)dr}{|E_{n\uparrow}-E_{n+1\downarrow}|},r<R_{0}
\end{equation}
\begin{equation}
h_{n\uparrow n+1 \downarrow}^{>}=-2\pi \alpha|\beta_{1}(\lambda)|^{2}\frac{
\int_{R_{0}}^{\infty}\lambda rK_{n}^{2}(\lambda r)dr}{|E_{n\uparrow}-E_{n+1\downarrow}|},r>R_{0}.
\end{equation}
The calculation of the counterpart $h_{n+1\downarrow n\uparrow}^{R}
=\frac{\langle n' \downarrow\vert h_{R}^{\dag} \vert n \uparrow\rangle}{|E_{n\uparrow}-E_{n+1\downarrow}|}$ follows the similar procedure, and the result is shown below.
\begin{equation}
\begin{aligned}
h_{n+1\downarrow n\uparrow}^{R} = &-2\pi\alpha[|\beta(\lambda)|^{2}
\int_{0}^{R_{0}}\lambda r I_{n+1}^{2}(\lambda r)dr\\
+&|\beta_{1}(\lambda)|^{2}
\int_{R_{0}}^{\infty}\lambda r K_{n+1}^{2}(\lambda r)dr]/|E_{n\uparrow}-E_{n+1\downarrow}|
\end{aligned}.
\end{equation}

When only the Rashba SOC is considered, although the matrix element $h_{n\uparrow n+1\downarrow}^{R}$ can be nonzero, there is no backscattering because of the constraint of energy conservation. This energy difference of initial and final states can be meet by phonon contribution. Thus, the electron-phonon interaction and Rashba SOC can give rise to an effective spin dephasing rate, and can further induce the inelastic backscattering.  We use the temperature-dependent spin dephasing rate to describe the inelastic backscattering \cite{PhysRevLett.83.1211} which can be written as
\begin{equation}
 1/\tau_{s}=8 \pi T \int_{0}^{\infty}d\omega \alpha_{s}^{2}F(\omega) \frac{\partial p(\omega)}{\partial T}.
\end{equation}
The spin-flip Eliashberg function $\alpha_{s}^{2}F$ reads \cite{Grimvall,PhysRevLett.83.1211}
\begin{equation}
\alpha_{s}^{2}F(\omega)=\frac{1}{\Delta}|g_{0}|^{2}|h_{n\sigma n'\sigma'}^{R}|^{2}\delta(\Delta-\hbar\omega)e^{-\frac{|\hbar\omega-\hbar\omega_0|^2}{2\xi^2}}
\end{equation}

\begin{figure}[H]
\centering
\includegraphics[scale=0.42]{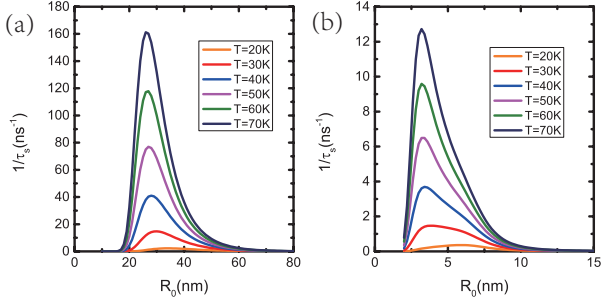}
\caption{(Color online)Spin dephasing relaxation time versus puddle size $R_0$ of (a) HgTe/CdTe QWs and (b) InAs/GaSb QWs. In (a), $mv^{2}=-24.8$ meV, $\hbar v=333.6$ meV.nm, $\omega_{0}=17$ meV\cite{Nanotech}; In (b), $mv^{2}=-7.8$ meV, $\hbar v=37$ meV.nm, $\omega_{0}=25$ meV\cite{PRB33.8889}; and we set the optical deformation potential $C_{op}=20$ eV , Rashba coupling constant $\alpha=70$ meV.nm\cite{PRBrashba}, phonon frequency width $\xi=3.5$ meV in both cases.} 
\label{fig:figure2}
\end{figure}

with $\Delta=|E_{n\sigma}-E_{n'\sigma'}|$ represents the energy difference,$\xi$ is the width of optical phonon energy. In eq. (24), $p(\omega)=[exp(\hbar\omega/k_{B}T)-1]^{-1}$
and $\hbar\omega$ represents the optical phonon energy. We assume the local electron-phonon scattering matrix as $g_{0}=\sqrt{\frac{\hbar C_{op}^{2}/a^{2}}{2\rho_{A}V \omega}}$, where $C_{op}$ is the optical deformation potential, $\rho_{A}$ is the atomic mass density and $a$ is the lattice constant \cite{PhysRevB.89.205103}.

In Figure 2, we show the spin dephasing relaxation time $1/\tau_{s}$ of HgTe/CdTe QWs and InAs/GaSb QWs as a function of radius. We consider Fermi energy locating in the band gap with $E_{F} = 0$. The linear $k$-order Rashba SOC accompanying by the electron-phonon interaction leads to inelastic backscattering in such a QSH puddle, which consequently provides an effective spin dephasing in the local helical edge states, although the system obeys the average TRS. The results are evaluated for puddles with size of tens-of-nanometer, where the helical edge states are located around the boundary $R_{0}$. Specifically for InAs/GaSb QWs the dephasing time largely depends on the puddle size, and$1/\tau_s$ is peaked with $1\sim14$ $ns^{-1}$ under low temperature, while for HgTe/CdTe QWs the dephasing time are much shorter. This result may explain the more robust edge transport in InAs/GaSb QWs. Next, for one certain temperature, $1/\tau_s$ varies with the puddle sizes and is peaked when the level spacing energy is around the phonon frequency $\omega_0$. When $\omega_0$ is matched, the spin dephasing effect is much stronger. Thus, the spin dephasing time is much shorter.The dephasing rate manifests as a non-monotonic function of puddles size at a given temperature. The reason is two-fold. Firstly, for small size puddles, the dephsing rate is small because the energy space between local edge states with different label becomes large, which can not be matched by the optical phonon energy. On the other hand, for puddles with large size, the edge states in puddles resemble the extended edge states, which also can hardly lead to spin dephasing with linear k-order Rashba SOC and normal dephasing\cite{PhysRevLett.108.086602}.

\begin{figure}[H]
\centering
\includegraphics[scale=0.4]{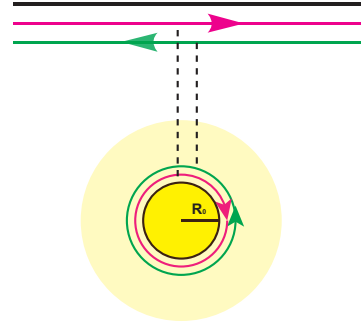}
\caption{(Color online)Tunneling between the extended helical edge states and local helical edge states in the puddles.} 
\label{fig:figure3}
\end{figure}

Finally, we discuss the the tunneling between the extended helical edge states at the boundary of sample and the local helical edge states surrounding the puddle(see Figure 3). In realistic two-dimension systems, puddles always exist with inhomogeneity due to long-range disorder or the external gate. The extended helical edge states at the boundary of sample can couple with the local helical edge states in the puddle several times. As a consequence, the electrons in extended helical edge states can also undergo the inelastic backscattering when spin dephasing happens in the puddles.

\section{Conclusion}

In conclusion, we consider the effective spin dephaing effect due to combination effect of $k$-linear Rashba spin-orbital effect and electron phonon interaction in two-dimension QSH QDs. We solve the energy spectrum of local helical bound states circulate around the boundary of QSH QDs, and calculate the spin dephasing relaxation time. Finally, we discuss the tunneling between the extended helical edge states and the local helical states in the QSH QD. We find that the effective spin relaxation time in InAs/GaSb system is much longer than that in HgTe/CdTe system, which is consistent with the more robust edge transport in InAs/GaSb QSH system.

\vspace*{2mm} \Acknowledgements{\bahao
This work was financially supported by NBRP of China (2015CB921102, 2012CB821402, and 2012CB921303) and NSF-China under Grants (Nos.11534001 an 11274364).}

\end{multicols}

\end{document}